# STT-SNN: A Spin-Transfer-Torque Based Soft-Limiting Non-Linear Neuron for Low-Power Artificial Neural Networks

Deliang Fan, Yong Shim, Anand Raghunathan, *Fellow, IEEE,* and Kaushik Roy, *Fellow, IEEE*

*Abstract*— Recent years have witnessed growing interest in the use of Artificial Neural Networks (ANNs) for vision, classification, and inference problems. An artificial neuron sums N weighted inputs and passes the result through a non-linear transfer function. Large-scale ANNs impose very high computing requirements for training and classification, leading to great interest in the use of post-CMOS devices to realize them in an energy efficient manner. In this paper, we propose a spin-transfer-torque (STT) device based on Domain Wall Motion (DWM) magnetic strip that can efficiently implement a Soft-limiting Non-linear Neuron (SNN) operating at ultra-low supply voltage and current. In contrast to previous spin-based neurons that can only realize hard-limiting transfer functions, the proposed STT-SNN displays a continuous resistance change with varying input current, and can therefore be employed to implement a soft-limiting neuron transfer function. Soft-limiting neurons are greatly preferred to hard-limiting ones due to their much improved modeling capacity, which leads to higher network accuracy and lower network complexity. We also present an ANN hardware design employing the proposed STT-SNNs and Memristor Crossbar Arrays (MCA) as synapses. The ultra-low voltage operation of the magneto metallic STT-SNN enables the programmable MCA-synapses, computing analog-domain weighted summation of input voltages, to also operate at ultra-low voltage. We modeled the STT-SNN using micro-magnetic simulation and evaluated them using an ANN for character recognition. Comparisons with analog and digital CMOS neurons show that STT-SNNs can achieve more than two orders of magnitude lower energy consumption.

## I. INTRODUCTION

SEVERAL neural network based computing models have been explored in recent years for realizing hardware that can perform human-like cognitive computing [1-6]. The fundamental computing units of such systems are the *neurons* that connect to each other and to external stimuli through programmable connections called *synapses* [1, 2]. The basic operation performed by an artificial neuron is computing a weighted sum of the *N* inputs and passing the result through a non-linear transfer function, expressed as follows:

$$Y = \varphi(\sum W_i \bullet IN_i - \theta) \quad (1)$$

where, $Y$ is the neuron output or *activation level*, $IN_i$ denotes the $i^{th}$ input, $W_i$ is the corresponding synapse weight, $\theta$ is the neuron threshold or bias and $\varphi$ is the neuron *transfer (activation) function*. Fig. 1b shows four representative neuron transfer functions. The step function is called *hard-limiting* transfer function because of the binary output states. The saturated linear, logistic sigmoid and hyperbolic tangent functions are *soft-limiting* transfer functions because of the continuous neuron output states [1, 2]. Large numbers of neurons can be connected in different network topologies to realize different neural network architectures [3-6]. For instance, cellular neural networks employ near neighbor connectivity [3], whereas, fully-connected feed-forward networks employ all-to-all connections between neurons in consecutive network layers or stages [4]. Several other network paradigms like Convolutional Neural Networks (CNN) [5], and Hierarchical Temporal Memory (HTM) [6] provide structured approaches to design large-scale networks. Irrespective of the network topology, neurons connect to each other in effect to communicate their probabilities (neuron activation levels) of being part of the final output [2]. The binary neuron output levels seriously hamper the possibility of neuron-to-neuron communication [2]. Soft-limiting neuron transfer functions are therefore preferred and greatly improve the neural network modeling capability while reducing network complexity. The reason behind this can be intuitively understood as follows. With hard-limiting functions, each neuron is required to decide whether it will be turned completely 'on' or completely 'off', which requires a step-like

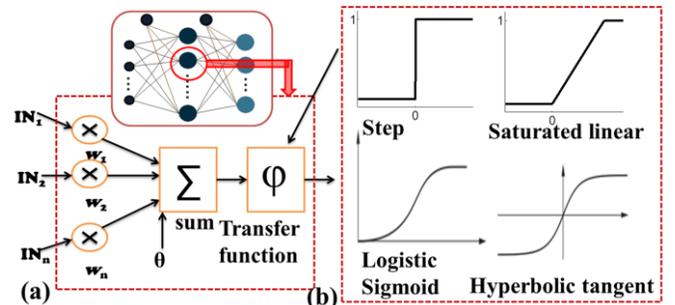

Fig. 1 (a) Artificial neuron: it takes weighted sum of n inputs and passes the result through an transfer/activation function (b) four representative transfer (activation) functions function. On the other hand, with soft-limiting functions, each neuron can be in any of a continuous range of activation levels between '0' and '1', allowing much more information to be communicated across neurons. Various functions that meet these requirements have been explored as artificial neuron transfer functions [1, 2, 28]. The optimal neuron transfer function is highly dependent on the dataset and network topology. In this work, we do not attempt to implement the optimal neuron transfer function, but rather propose an energy-efficient spin-torque based device that can implement a continuous non-linear function, which can be used as a soft-limiting artificial neuron transfer function.

The energy efficiency, performance, and integration density of ANN hardware is governed by the design of the fundamental computing units that realize neurons and synapses. Prior works in this field involved the development of circuits for neurons and synapses using CMOS, and in general, employed large numbers of transistors and required high power consumption [7, 8]. Therefore, in order to translate the ANN algorithmic models into powerful, yet energy-efficient cognitive computing hardware, computing devices beyond CMOS are being explored. Recent experiments on spin-torque devices have demonstrated high speed switching of nano-magnets with small currents [9-12]. Such magneto-metallic devices can operate at ultra-low terminal voltages and can implement current-mode summation and non-linear operations required by an artificial neuron. We previously proposed the application of spin-torque neurons based on lateral spin value (LSV) and domain wall motion (DWM) magnet for designing ultra-low power neural networks [13-15]. However, all of the previously proposed spin-neurons implement the hard-limiting step-function, which leads to larger network size, and simply cannot provide adequate modeling accuracy for complex classification problems.

In this paper, we propose a Spin-Transfer-Torque based Soft-limiting Non-linear Neuron (STT-SNN) having an output which is a rational function of the total incoming synapse currents, leading to compact network size and ultra-low power consumption. Instead of binary output states, our proposed STT-SNN can have continuous output voltages. We also present an ANN hardware design employing deep-triode current source (DTCS) transistors as interfacing circuits and memristor crossbar arrays (MCA) as synapses. The fact that STT-SNNs operate at ultra-low voltages enables the programmable MCA synapses, computing analog domain weighted summation of input voltages, to also operate at ultra-low voltage for low overall energy consumption. Comparison with state-of-art digital/analog CMOS neurons shows that the proposed spin based neuron can achieve more than two orders of magnitude lower energy.

The rest of the paper is organized as follows. Previous work on hard-limiting spin-neurons is briefly introduced in section II. Section III presents the proposed device structure and circuit model for the proposed soft-limiting spin based neuron. The use of MCA as synapses is described in section IV. Section V presents the overall hardware implementation of ANNs using the proposed STT-SNNs. Section VI discusses the performance of the proposed ANN design for a benchmark application (character recognition) and its comparison with other recent neuron implementations. Section VII summarizes and concludes the paper.

## II. PREVIOUS WORKS ON HARD-LIMITING SPIN-NEURONS

Previously, we proposed the application of hard-limiting spin-neurons based on lateral spin valves (LSV) [13], as well as domain wall motion (DWM) magnets [14, 15] for designing ultra-low power artificial neural networks.

### A. Bipolar Lateral Spin Valve Neuron

Fig. 2a shows the device structure of a bipolar spin-neuron based on LSV. It consists of a high polarization (P) input magnet m2-m4 acting as a spin injector and a low polarity output magnet m1 forming a Magnetic Tunnel Junction (MTJ) based read port with a fixed magnet. The two anti-parallel, stable polarization states of a magnet (m2 and m3) lie along its easy axis. The direction orthogonal to the easy axis is an unstable polarization state for the magnet and is referred as its hard-axis. Charge current injected into the channel through m2

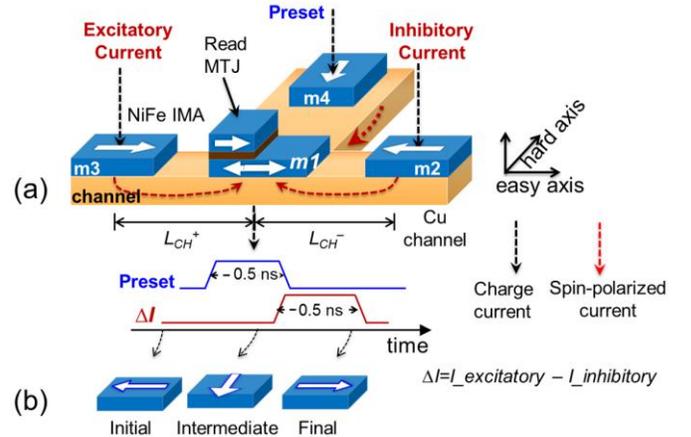

Fig. 2 (a) Spin-neuron based on LSV with two complementary inputs (b) spin neuron states and m3 gets spin polarized according to the corresponding polarity of magnets. Spin polarized charge current is modeled as a four component quantity, one charge component $I_c$, and three spin components ($I_x, I_y, I_z$) [13]. Each of these two anti-parallel spin polarized currents exerts a spin transfer torque (STT) on m1, switching the spin polarization of m1 along the easy axis. The preset magnet m4 shown in Fig. 2a, however, has its easy-axis orthogonal to that of m1, and is used to implement current mode Bennett clocking [13]. A current pulse input through m4, presets the output magnet, m1, along its hard axis (Fig. 2b). The excitatory and inhibitory synapse current pulses are received through the magnets, m3 and m2. After removal of the preset pulse, m1 switches back to its easy axis, which is parallel to that of m2 and m3. The final spin polarity of m1 depends upon the difference-$\Delta I$ between the spin polarized charge current inputs through m3 and m2, corresponding to excitatory synapse current and inhibitory synapse current. Hard axis, being an unstable state for m1, even a small value of $\Delta I$ effects deterministic easy-axis restoration. Note that, the lower limit on the magnitude of $\Delta I$ (hence, on current per-input for the neuron), for deterministic switching, is imposed by the thermal noise in the output magnet, and, imprecision in Bennett Clocking. The effects of these non-idealities have been included in device simulation [13]. The read MTJ effective resistance is larger when the spin polarity of m1 is anti-parallel to the fixed magnet and vice versa. A dynamic CMOS latch is used to sense the resistance of read MTJ. Thus, the thresholding operation (step function) of the synapse currents can be implemented efficiently using this LSV based 'spin-neuron'.

*B. Unipolar Domain Wall Motion Neuron*

A domain wall motion (DWM) based magnetic strip constitutes of multiple *nano-magnet* domains (d1, d2) separated by non-magnetic region called *domain wall* (DW) as shown in Fig. 3a. DW can be moved along a magnetic nano-strip using current injection along the DWM strip [9-11]. Hence, the spin polarity of the DWM strip at a given location can be switched, depending upon the polarity of its adjacent domains and direction of current flow. Fig. 3b shows the DW is moved to left by the spin polarized electrons from d2. Recent experiments have achieved DW depinning critical current

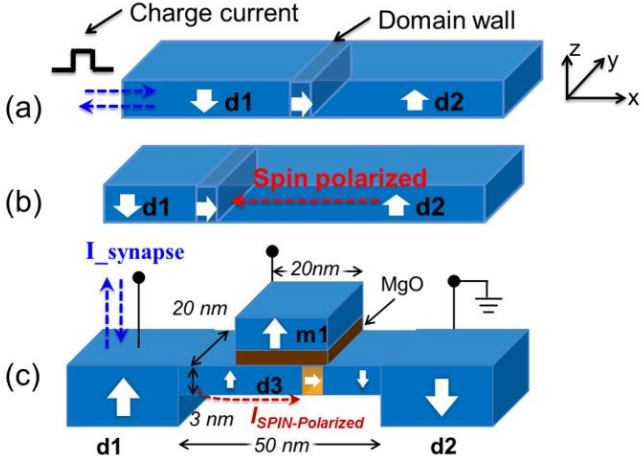

Fig.3 (a) A domain wall magnet with two domains, (b) domain wall is pushed to left by the spin polarized electrons (c) device structure for domain wall neuron (DWN).

density of ~$6\times10^{11}$A/m$^2$ and ~60m/s DW moving velocity for 20nm-wide DWM strips [9].

The previously proposed Domain Wall Neuron (DWN) device structure is shown in Fig 3c. It constitutes of a thin and short ($3\times20\times50$ nm$^3$) nano-magnet domain, d3 (the 'free domain') connecting two anti-parallel nano-magnet domains of fixed polarity, d1 and d2. Domain-1 forms the input port, whereas, d2 is grounded. The total synapse currents are injected through d1. Spin polarity of the free domain (d3) can be written parallel to d1 by the spin-polarized electrons from d1 to d2 and vice-versa. Apart from device scaling, the use of lower anisotropy barrier for the magnetic material can be effective in lowering the switching threshold for computing applications. A magnetic tunnel junction (MTJ), formed between a fixed polarity magnet (m1) and d3 is used to read the state of d3. The effective resistance of the MTJ is smaller when m1 and d3 have the same spin polarity and vice-versa. We employ a dynamic CMOS latch to detect the MTJ state. Thus, the DWN can detect the polarity of the current flow at its input node. It acts as a low power and compact current comparator that can be employed as energy efficient current mode hard limiting step function artificial neuron. Note that, this current can be further reduced by lowering the energy barrier or applying spin-orbital coupling [15].

The previously proposed spin based neurons can achieve energy efficient step function as transfer function for artificial neurons. However, as we discussed earlier, soft-limiting neuron transfer functions are preferred because of their improved modeling capacity of ANNs, leading to compact ANN design for the same application. In the next section, we propose a spin-torque device that can implement a soft-limiting non-linear neuron transfer function.

### III. PROPOSED SPIN-TRANSFER-TORQUE BASED SOFT-LIMITING NON-LINEAR NEURON

In this section, we describe the device structure and operation of the proposed soft-limiting neuron. The CMOS circuits employed to interface to the neuron are also discussed.

The proposed Spin-Transfer-Torque based Soft-limiting Non-linear Neuron (STT-SNN) is based on a composite device structure consisting of a DWM magnetic strip and a magnetic tunnel junction (MTJ) as shown in Fig. 4a. The MTJ consists of two ferromagnetic layers with an MgO barrier sandwiched between them. The 'free' ferromagnetic layer (d4) connects laterally to two anti-parallel fixed domains-d1 and d2 [12, 21]. The larger thickness at the edges of the free layer is used to stabilize the DW at an intermediate position within the free layer [12]. In general, the application of current induced domain wall motion faces the problem of stable control of domain walls. It comes from many reasons, such as DW structural change, bidirectional displacements, stochastic nature of DWM, thermal effect of Joule heating and the local pinning effect [31-35]. These problems can be largely solved by reducing the critical current density required to de-pin the domain wall from a pinning site. A small DWM critical current density in the range of $10^{11}$A/m$^2$ was demonstrated experimentally in a scaled magnetic nanostrip with Perpendicular Magnetic Anisotropy (PMA) [9]. The reason why PMA device has a smaller DWM critical current density compared with In-plane Magnetic Anisotropy (IMA) device can be explained as follows. In the magnetic nanostrip, when the current is injected through a fixed domain, it becomes spin-polarized and exerts a torque on the DW. This torque induces the rotation of magnetization to the hard-axis direction, resulting in the pinning force. If the current density is above a certain threshold, the Spin-Transfer-Torque (STT) can overcome this pinning force, leading to steady DWM. Thus, the critical current density can be lowered by increasing the STT (narrower domain wall) or decreasing the pinning force (lower hard-axis anisotropy). In summary, the critical current density-$j_{th} \propto K_{h.a.}\Delta$, where $K_{h.a.}$ is hard-axis anisotropy and $\Delta$ is the domain wall length [31-35]. The hard-axis anisotropy of a PMA device reduces with lower device thickness and becomes much smaller than that of an IMA device. Moreover, the DW length in a PMA device is in general smaller than that in an IMA device. Therefore, a scaled PMA magnetic nanostrip is used in our work to achieve lower critical current density to induce steady DWM. The free layer dimensions are $2\times20\times100$nm$^3$ as shown in Fig. 4a. A *Neel* type DW is formed because of the small strip width (20nm) [9]. The DW length $L_{DW}=\pi\sqrt{(A_{ex}/Ku)}=$ ~17nm based on our device parameters listed in table-I.

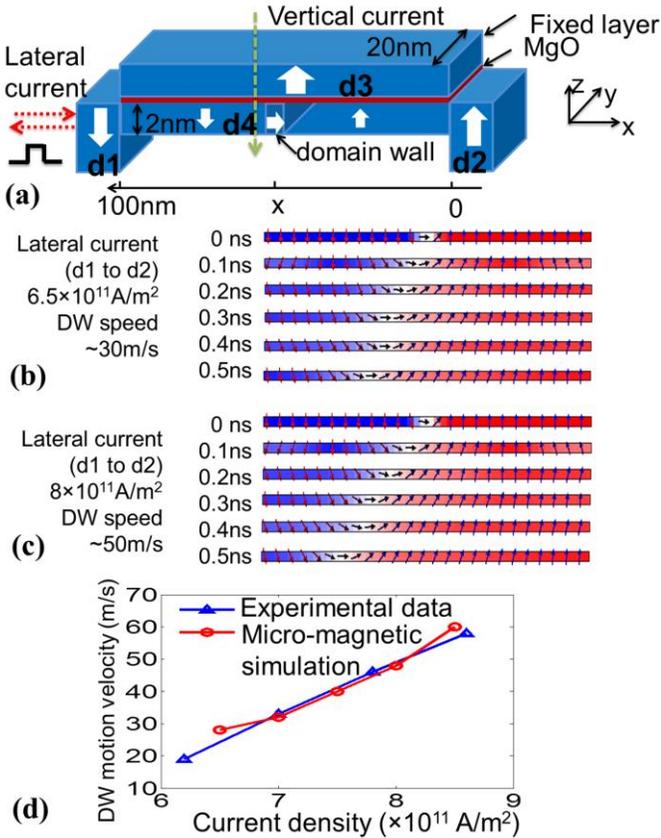

Fig. 4 (a) The proposed STT-SNN device structure, (b) the micro-magnetic simulation of free layer DW motion when the injected lateral current density is $6.5\times10^{11}$ A/m$^2$ and (c) $8\times10^{11}$ A/m$^2$, (d) simulated DW motion velocity vs. current density, showing a good match with experimental data reported in [9]

The proposed STT-SNN device can be treated as a four terminal device with lateral and vertical current paths. For the lateral path (d1 to d2, $\pm$x direction), d1 forms the input programming port, assuming d2 is supplied with a constant voltage. The domain wall can be moved along the free layer depending on the lateral current pulse magnitude, direction and duration [9-11], leading to a continuous resistance change of the MTJ in the vertical direction. The transient micro-magnetic simulation plot of the free layer using *mumax*$^3$ [16] is shown in Fig. 4b&c, where a 0.5ns current pulse with magnitude of $6.5\times10^{11}$A/m$^2$ and $8\times10^{11}$A/m$^2$ are applied from d1 to d2. It can be seen that the domain wall moves to the left (along the direction of electron flow) with a different speed. The device parameters used in the simulation are listed in table-I. We benchmarked the micro-magnetic simulation with the experimental data in [9] (the same nano-strip width of 20nm is fabricated in the reference) and it shows a good match as shown in Fig. 4d. A relatively high *Ku* (i.e. high energy barrier) is preferred in the memory application for the sake of good thermal stability [9]. In the computing applications, a lower energy barrier can be used to reduce the critical current density to de-pin the DW, which leads to lower energy consumption.

The vertical path (from d3 to d4, $\pm$z direction) is used for sensing the position of DW in terms of MTJ vertical resistance. MTJ resistance is a function of voltage, tunneling oxide thickness ($t_{ox}$) and the angle between free layer and pinned layer magnetizations. The atomistic level simulation framework based on Non-Equilibrium Green's Function (NEGF) formalism [18] can be used to evaluate the MTJ resistance, which includes the device variation and thermal fluctuation. A behavioral model based on statistical characteristics of the device is used in SPICE simulation to assess the system functionality. It models the device as three parallel MTJs with variable resistance depending on DW positions (Fig. 6a):

$$R_L = RA_{AP} / (W \bullet (L - x - 0.5 L_{DW})) \quad (2)$$

$$R_R = RA_P / (W \bullet (x - 0.5 L_{DW})) \quad (3)$$

$$R_{DW} = RA_{DW} / (W \bullet L_{DW}) \quad (4)$$

$$R_{neuron} = R_L // R_{DW} // R_R = \frac{A}{B \bullet x + C} \quad (5)$$

$$A = RA_{AP} \bullet RA_P \bullet RA_{DW} \quad (6)$$

$$B = (RA_{AP} - RA_P) RA_{DW} \bullet W \quad (7)$$

$$C = RA_P \bullet RA_{DW} \bullet W \bullet L + \\ (RA_{AP} \bullet RA_P - 0.5 RA_P \bullet RA_{DW} - 0.5 RA_{AP} \bullet RA_{DW}) W \bullet L_{DW} \quad (8)$$

where, $R_{neuron}$, $R_L$, $R_{DW}$ and $R_R$ are respectively the vertical resistance of STT-SNN, left anti-parallel, domain wall and right parallel equivalent MTJ resistances; $x$ is DW position (middle point), $L$ is the length of free layer (100nm), $W$ is the width of free layer, $RA_{AP}$, $RA_{DW}$ and $RA_P$ are respectively MTJ resistance-area product for anti-parallel, DW and parallel configurations. The values we used in the simulations are: $RA_{AP}$=5Ω•μm$^2$, $RA_{DW}$=~3.5Ω•μm$^2$, $RA_P$=2Ω•μm$^2$ [12, 18]. Note, this model is used for SPICE simulation in sensing the neuron state. DW position (x) is a function of total input currents, modeled using micro-magnetic simulation as described earlier.

The interface circuit of STT-SNN is shown in Fig. 5a. It works in three phases – programming, sensing and reset phase. In the programming phase, the lateral programming current (total synapse current) programs DW position along the free layer. Then, for the sensing phase, a voltage divider circuit is used to sense the STT-SNN state. The reference MTJ voltage is treated as neuron output voltage which will be transmitted through 'axon' to its fan-out neurons (axon circuit will be explained in section V). For maximum power efficiency and the isolation of two paths, different phases should be separately powered. The clocked power supplies called *pClocks* can be used (as shown in Fig. 5b). When in the programming and the reset phases, *PclkB+* and *PclkB-* are in *floating state*, while *PclkA* provides a constant voltage *V* to d2, enabling the lateral programming path. When it is in the sensing phase, *PclkA* and the input terminal (d1) are in the floating state. Meanwhile, *PclkB+* and *PclkB-* supply 50mV and -50mV, respectively (choice of sensing voltage will be explained later). The clocked power supply can be implemented using widely used power gating technique [20]. Finally, a reset current pulse (-50μA, 1ns) is applied to the STT-SNN free layer to set the DW

TABLE I
DEVICE PARAMETERS USED IN SIMULATION

| Symbol | Quantity | Values |
|---|---|---|
| α | damping coefficient | 0.02 |
| Ku | uniaxial anisotropy constant | $3.5\times10^5$ J/m$^3$ |
| Ms | saturation magnetization | $6.8\times10^5$ A/m |
| $A_{ex}$ | exchange stiffness | $1.1\times10^{-11}$ J/m |
| P | polarization | 0.6 |

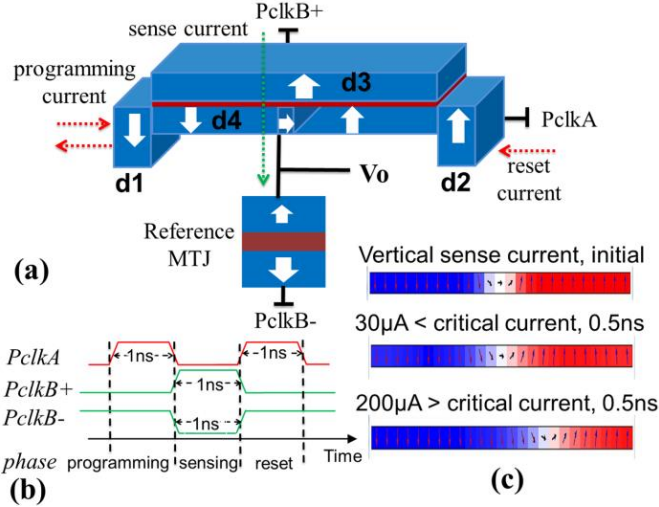

Fig. 5 (a) The programming and sensing circuit of the proposed STT-SNN, (b) the clocked power supply waveforms, (c) the micro-magnetic simulation of STT-SNN free layer with different vertical sense currents.

location in the rightmost corner, ready for the next computation cycle.

The authors in [12] have experimentally shown that the vertical current may also shift DW when the current density is above a critical value because of the out-of-plane ('field-like') spin transfer torque. DW position displacement is what we want to avoid in sensing the STT-SNN resistance. Note, the DW position essentially indicates the state of the neuron. Based on the micro-magnetic simulation for vertical current injection, the vertical critical current density to de-pin the DW was found to be ~$5\times10^{10}$A/m$^2$ [12], corresponding to a critical current of ~100μA. The reference MTJ resistance in Fig. 5a is 2.5kΩ and the STT-SNN resistance is in the range of ~1kΩ to ~2.5kΩ depending on the DW position. Therefore, the largest allowed voltage difference between *PclkB+* and *PclkB-* is ~350mV. In order to keep a good amount of sensing margin, *PclkB+* and *PclkB-* are set to be 50mV and -50mV, respectively, which corresponds to a maximum of 30μA vertical sensing current. From the micro-magnetic simulation shown in Fig. 5c, DW position is stable when the vertical sensing current is 30μA.

Based on the compact STT-SNN model, the output voltage in Fig. 5a) can be computed as:

$$V_O = V_s \frac{R_{ref}}{R_{ref} + R_{neuron}}$$
$$= V_s (1 - \frac{A}{R_{ref} \cdot B \cdot x + R_{ref} \cdot C + A}) \quad (9)$$

where, $V_s$ is the voltage difference between *PclkB+* and *PclkB-* (100mV), $R_{ref}$ is the reference MTJ resistance, $x$ is the domain wall location, $A, B, C$ are the constants of equations 6-8. It can be observed that the output voltage is a *rational function* of DW positions (0<$x$<100nm). Note, rational function is defined as the ratio of two polynomials (two linear functions with the same slope in our case). '$x$' is a function of the total lateral programming current as described earlier. Fig. 6b shows the STT-SNN resistance vs. DW position. It can be seen that the STT-SNN resistance can be adjusted in a continuous range of values based on the DW position, enabling continuous output voltages as shown in Fig. 6c. Based on the micro-magnetic simulation of DW motion velocity dependence on the injected current density shown in Fig. 4d, the neuron output voltage vs. programming current (assuming 1ns clock cycle) is

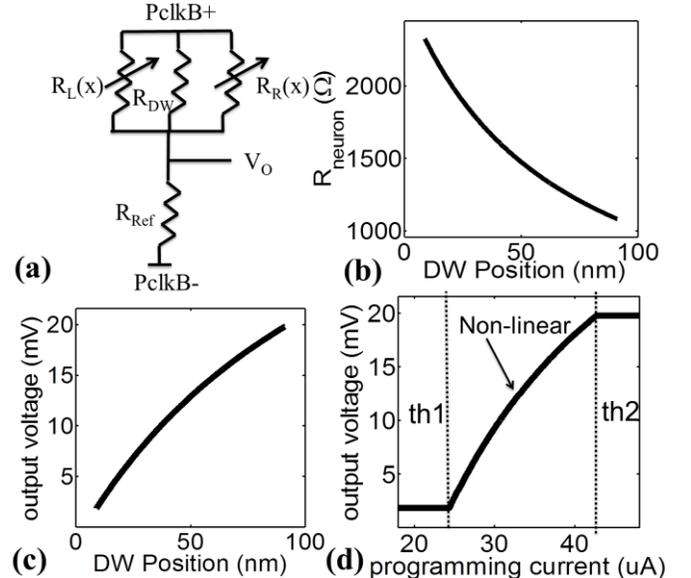

Fig. 6 (a) Behavioral SNSN SPICE model, (b) SNSN resistance vs. DW positions, (c) output voltage vs. DW positions, (d) output voltage vs. programming current. Note, the positive current direction is defined from d1 to d2 as shown in fig. 5a. Clock cycle is 1ns.

plotted in Fig. 6d. The positive current direction is defined as from 'd1' to 'd2' as shown in Fig. 5a. Note that, the programming current here is the total synapse current (weighted sum of inputs in ANN model). If the programming current is smaller than the DW depinning critical current (*th1*), DW is stable at the initial position and the output voltage is minimum. When the programming current is larger than '*th2*', DW will be pushed to the other end and the output voltage saturates to the maximum. '*th2*' can be defined as the minimum current to push the domain wall from one end to the other end with 1ns clock cycle. This two threshold currents (*th1* and *th2*) can be tuned by proper device dimensions and material parameters to adapt different ANN designs. From the above discussions it is clear that the proposed device can be used to implement the low current, high speed, soft-limiting non-linear function of an artificial neuron. Next, we will show that the weighted summation of inputs can be efficiently implemented by MCA-synapse.

## IV. MEMRISTOR CROSSBAR ARRAY SYNAPSES

The two-terminal synapse bears striking resemblance to memristor whose conductance can be precisely modulated by charge or flux through it [22]. In the ANN model shown in Fig. 1a, the inputs go through the associated synapses (multiplied by weights) and are summed up as input to the neuron transfer function. This operation can be implemented efficiently using a memristor crossbar array (MCA) shown in Fig. 7a [23]. It constitutes of memristors (e.g. Ag-Si) with conductivity $g_{ij}$, interconnecting two sets of metal bars ($i^{th}$ horizontal bar and $j^{th}$ in-plane bar). Input voltages $V_i$ can be applied to horizontal bars. Assuming the outward ends of the in-plane bars grounded, the current going through the interconnected memristor is $V_i \cdot g_{ij}$. Therefore, the total current coming out of the $j^{th}$ in-plane bar can be visualized as the dot product of the inputs $V_i$ and the memristor conductance-$g_{ij}$ (Fig. 7a), expressed as $\Sigma_i V_i \cdot g_{ij}$. The MCA stores the synapse weights in terms of memristor conductance, and evaluates the weighted sum of the inputs required for the ANN.

High precision, multi-level write techniques for isolated memristors have been proposed and demonstrated in literature that can achieve more than 8-bit write accuracy [24]. In our work, 5 bit accuracy was chosen for demonstrating system functionality. Note, lower synapse weight resolution can be used by increasing the number of neurons. It is a trade-off between the resolution of the weights and the number of neurons. Even binary weight configuration can be used, however, that would require much more neurons. In a crossbar array consisting of large number of memristors, write voltage applied across two cross-connected bars for programming the interconnecting memristor can result in *sneak current paths* through neighboring devices [25]. This disturbs the state of unselected memristors. To overcome the sneak path problem, application of access transistors and diodes have been

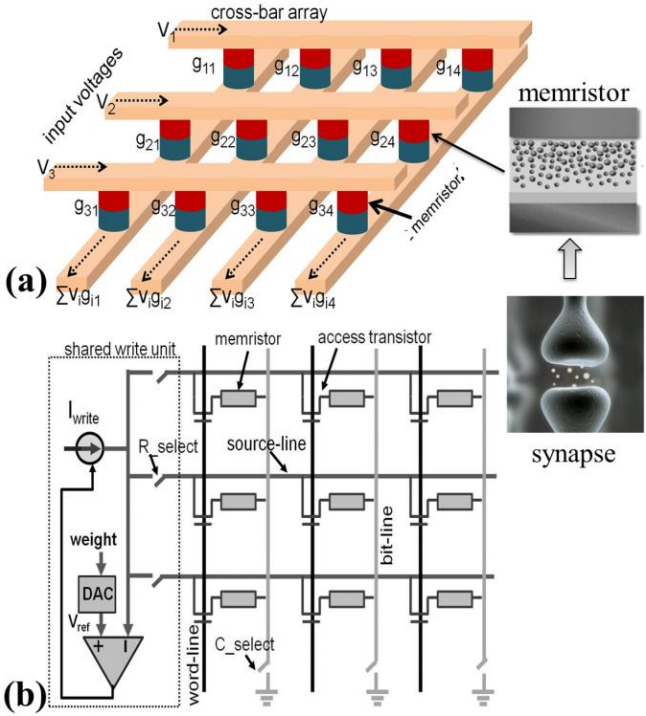

Fig. 7 (a) Memristor crossbar array used for evaluating the weighted sum of inputs for ANN (b) peripheral circuits for MCA programming [14]

proposed in literature [25] that facilitate selective and disturb free write operations. A multi-bit memristor array-level programming scheme employing adjustable pulse width is shown in Fig. 7b [14]. In this scheme, when programming one specific memristor cell in the array, the corresponding set of the word line, source line and bit line will be selected. During the writing operation, a constant current will be injected into the selected cell and the voltage developed on the source line is compared with a comparator threshold. A digital to analog converter (DAC) is used to set the threshold proportional to the target resistance. As soon as the accessed memristor is programmed to the target value, the current source is disconnected [14]. More precise tuning of memristor value can be achieved by applying a lower value of write current resulting in slower ramp in the resistance value. The memristive devices (including Ag-Si) do exhibit a finite write threshold for an applied current/voltage, below which there is negligible change in resistance [26]. Since the application of spin based neuron facilitates ultra-low voltage (and hence low current) operation of the memristors for computing, the state of memristor in the MCA will not be disturbed during read operations.

## V. ANN Hardware Using STT-SNN And MCA

In this subsection, we describe our proposed ultra-low power ANN hardware design combing MCA synapses and STT-SNN, showing one to one similarity to biological neural network.

In a biological neural network, 'axons' are used to transmit electrical-chemical signal between neurons [1, 2]. In our proposed ANN hardware (Fig. 8), a deep triode current source (DTCS) transistor is used to act as an 'axon' interconnecting the previous stage neuron output (voltage) with MCA synapses. As shown in Fig. 9a, the drain to source voltage of DTCS transistor is of the order of few tens of millivolts and it operates in the '*deep-triode*' region where the drain current $I_{ds}$ is linearly proportional to $Vdd\text{-}V_T\text{-}V_g$, where $V_T$ is the threshold voltage and $V_g$ is the gate voltage. Moreover, the maximum $I_{ds}$ can be tuned by the width of the transistor and $V_{ds}$ as shown in Fig. 9a. Therefore, DTCS transistor can be used to transmit the neuron output voltage into synapse current similar to axon [13]. Fig. 8 shows the spin-CMOS hybrid ANN (one layer) hardware design using DTCS-axon, MCA-synapses and STT-SNN, which shows one to one similarity to biological neural network. The $i^{th}$ input to the MCA synapses may connect to the $j^{th}$ STT-SNN with either positive, negative or zero weight. This is achieved by programming either $g_{ij+}$ or $g_{ij-}$ to the corresponding weight. For zero weight (i.e. no connectivity), both $g_{ij+}$ and $g_{ij-}$ are driven to high resistance '*off*' state. The input signal to MCA synapses is received through DTCS transistors with source terminals connected to a potential $V+\Delta V$ (for positive weights) and to $V-\Delta V$ (for negative weights), where $\Delta V$ can be ~50mV. Ignoring the parasitic resistance of the metal crossbar (for small scale network size), the current going through one synapse can thus be written as $I_{in}(i) \cdot g_{ij}/g_{TR}$, where $I_{in}(i)$ is the current supplied by the $i^{th}$ DTCS transistor, $g_{ij}$ is the synapse weight dependent conductance of the $i^{th}$ input to the $j^{th}$ neuron and $g_{TR}$ is the total conductance (of all the Ag-Si memristors) connected to a horizontal bar (dummy memristors are added such that $g_{TR}$ is equal for all horizontal bars). As a result, the current coming out of each MCA in-plane bar is the total current going into the connected STT-SNN, and can be expressed as $\Sigma I_{in}(i) \cdot (g_{ij+} - g_{ij-})/g_{TR}$, where $I_{in}(i)$ is linearly proportional to the input voltage. The total synapse current determines the STT-SNN output voltage according to the soft-limiting non-linear transfer function shown in Fig. 6d.

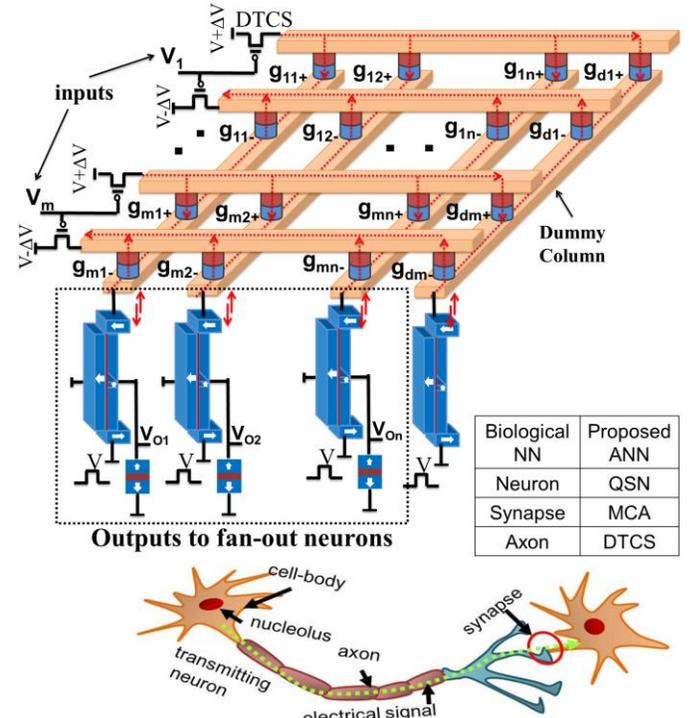

Fig. 8 The proposed ANN hardware design using DTCS-axon, MCA-synapse, and STT-SNN

The linearity and source-to-drain current range of DTCS is affected by the fluctuation in drain voltage. Lowest possible range of values for the memristor resistances, hence the highest $g_{TR}$ would largely overcome the non-linearity of DTCS output current as shown in Fig. 9b. Note that voltage drop in the metal lines due to parasitic resistances is small compared with memristor resistances. The other design parameters like the synapse weight resolution, neuron transfer function thresholds etc., were determined based on the simulation of the MCA model [26] and neural network training to ensure the

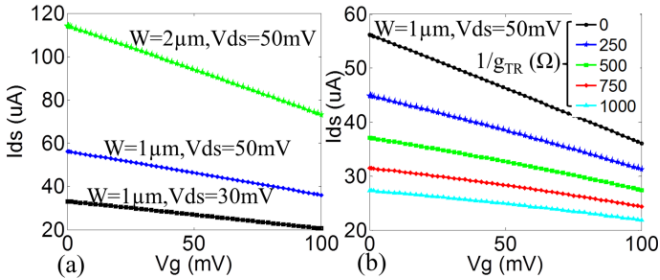

Fig. 9 (a) DTCS Ids vs. Vg for different width and Vds (b) non-linearity characteristics of DTCS transistor due to drain terminal memristor load implemented ANN accuracy. The required range of current output from DTCS is determined by the network size, weight resolution of synapses, $g_{TR}$ and neuron threshold. The current range can be obtained using different combination of DTCS sizing and the terminal voltage-$\Delta V$ as shown in Fig. 9a. For a required amount of DTCS current, it is desirable to push $\Delta V$ to the minimum possible value, in order to reduce the static power consumption in the MCA. This would imply exploiting the low voltage operation of the STT-SNN to the maximum possible extent. The minimum value of $\Delta V$ is limited mainly by the non-linearity of DTCS that degrades the output neuron *detection margin* (difference between the highest output to the second highest output) and hence, the matching accuracy. For the benchmark we describe in the next section, $\Delta V$ of 50mV (with regulated DC supply of 1mV prevision [30]) was found to be enough to preserve the matching accuracy close to the ideal case. The proposed scheme effectively biases the MCA-synapses across a small terminal voltage $\Delta V$ (between two DC supplies: $V+\Delta V$ and $V$), thereby ensures that the MCA computes the weighted sum of the inputs at low power.

## VI. APPLICATION & PERFORMANCE RESULTS

In this section, we describe the performance of the proposed ANN design for a benchmark application (character recognition) and its comparison with other CMOS and spin based neuron designs.

We simulated character (English alphabets) recognition as a benchmark application using the proposed ANN design. The CMOS peripheral circuits are simulated using IBM 45nm SOI technology. The overall process for character recognition can be divided into two steps, namely, edge extraction and pattern matching. Note that the edge extraction and ANN training are performed offline. Each alphabet feature vector is composed of 64 components extracted from four directions: horizontal, vertical and $\pm 45^0$ [13] (Fig. 10). Each 64-component feature vector is one test vector to a pre-trained feed-forward ANN composed of hidden layer and output layer as shown in Fig.

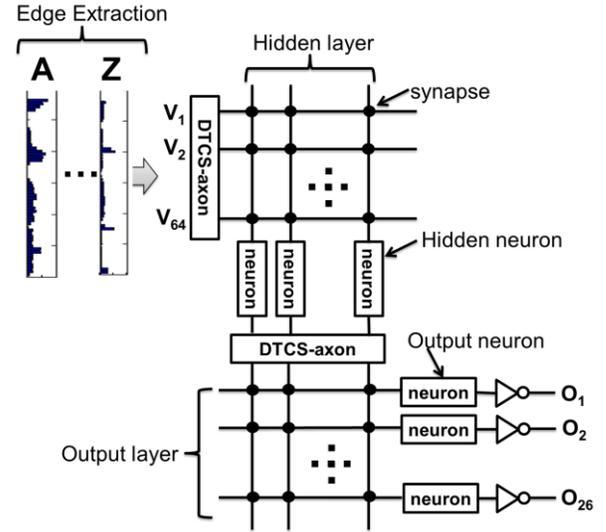

Fig. 10 Alphabet feature vectors and two-layer feed-forward ANN architecture. Note, the hardware implementation of each layer can be seen in fig. 8

TABLE II
NUMBER OF NEURONS FOR DIFFERENT NEURON TRANSFER FUNCTIONS

| Transfer functions | Hard-limiting | Soft-limiting | | |
| --- | --- | --- | --- | --- |
| | step | Sat_linear | Sigmoid | STT-SNN |
| # of hidden neuron | 24 | 9 | 4 | 5 |
| # of output neuron | 26 | 26 | 26 | 26 |

10. Table-II shows the MATLAB neural network training results using four different neuron transfer functions for the same benchmark and recognition accuracy. It can be seen that the hard-limiting step-function requires much more hidden neurons than the other soft-limiting neurons. It is mainly because the soft-limiting neuron, with a continuous output, has a much larger modeling capacity. Thus, as a soft-limiting neuron model, our proposed STT-SNN can achieve a more compact network size compared to hard-limiting neurons. The mapped hidden layer area can be seen in Fig. 12b. For all cases, the number of output neurons is the same, since each output neuron corresponds to one alphabet.

In the ANN architecture as shown in Fig. 10, DTCS-axons in the first (hidden) layer take the analog voltage inputs proportional to input feature vectors and convert them to current going through the MCA-synapses. In all, 64×2 DTCS-axons (positive and negative weights) are required and the MCA (synapse matrix) size is 128×6 (5 hidden neurons and one dummy column). The output layer contains 5×2 DTCS-axons and the MCA size is 10×27 (26 output neurons and one dummy column). Note that, a Gaussian distributed random noise (σ=5%) was added to each memristor conductance value in our simulations to model variations. The simulation results are shown in Fig. 11a. The figure shows the normalized output neuron voltages for 26 test alphabets. Pixel ($i$, $j$) indicates the $i^{th}$ output neuron voltage when the input is the $j^{th}$ alphabet. During the supervised training of the ANN, the 26 output neurons (O1 to O26) are assigned to indicate 26 alphabets ('A' to 'Z') respectively. Thus, for each test alphabet (each row in Fig. 11a), the diagonal value-($i$, $i$) should be the maximum to

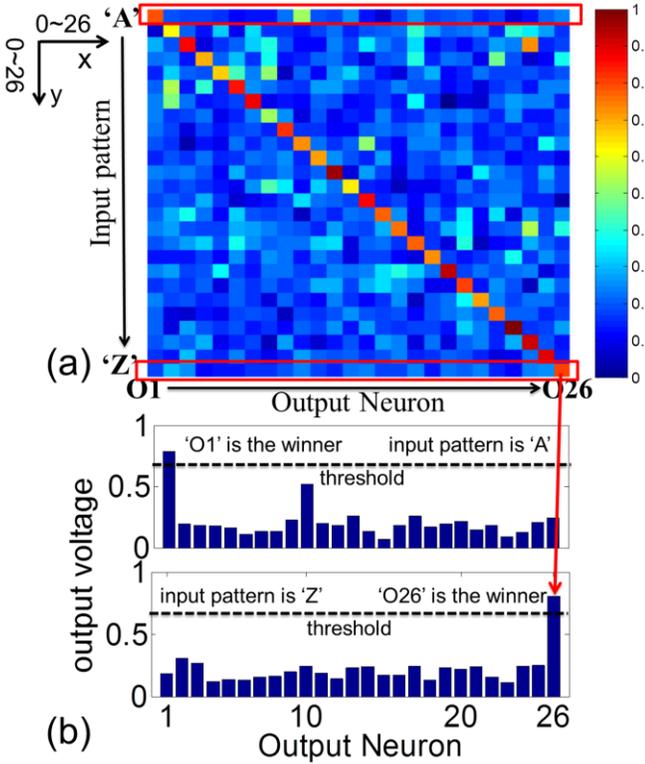

Fig. 11 (a) Normalized 26 output neurons' voltage for 26 test input patterns. Note that, pixel ($i$, $j$) indicates $i^{th}$ output neuron voltage for $j^{th}$ input pattern. (b) The 26 output neurons' voltages when the input patterns are 'A' and 'Z' indicate a correct match. The first ('A') and last row ('Z') voltage values are separately plotted in Fig. 11b. It can be seen that, when the input pattern is 'A', output neuron-'O1' is the winner. In the case that 'Z' is the input pattern, output neuron-'O26' is the winner. For the output winner detection, a simple Winner Take All (WTA) circuit described in [29] can be employed. Based on SPICE simulation for this simple alphabet benchmark, we found the voltage difference between the winner and other output neurons is sufficiently large (Fig. 11a). Thus, we attached an inverter to each output neuron to sense the output. Only the winner output bit is '0', while the others are '1's.

The energy consumption of a single STT-SNN has three components: programming, sensing and reset energy. For an average of ~40μA of lateral current flowing across the STT-SNN free layer (the total current out of one MCA column/row) and an effective lateral resistance of ~300Ω, the programming energy is evaluated to be ~0.5fJ for 1ns clock cycle time. The second component (sensing energy) can be ascribed to the MTJ-based read operation as described in section III. A read current of ~25 μA (~20% of DW depinning vertical critical current) would lead to ~2.5fJ energy consumption for 1ns read speed. Note that, the sensing current and sensing energy can be reduced by increasing the MTJ MgO thickness (hence, the resistance-area product of MTJ [18]). For the reset energy, a 50μA-1ns current pulse is used in our simulation, leading to ~0.75fJ reset energy. Thus, the total energy dissipation of one single STT-SNN is ~3.75fJ. Note that, each phase delay is set to be the same (1ns) to make it easy for pipelining the design. We compare the proposed STT-SNN energy with other recent artificial neuron implementations in Fig. 12a. Compared with CMOS analog and digital neurons in [15, 27], STT-SNN leads to the possibility of more than two orders of magnitude lower energy dissipation. The LSV-based spin-neuron (step function) is around one order of magnitude larger than STT-SNN because of the large hard-axis preset energy [13]. The reasons why the energy consumption of DWM spin-neuron (step function) [15] is smaller than that of STT-SNN is mainly due to $1)$ spin-orbital coupling is employed to increase the DW velocity; $2)$ a smaller sense current is used; $3)$ it implements a step function with hysteresis and no reset operation is required.

Apart from the ultra-low energy consumption, the soft-limiting functionality of STT-SNN also leads to reduced number of hidden neurons, and hence smaller hidden layer area for the same benchmark [1, 2]. In our SPICE simulation, the

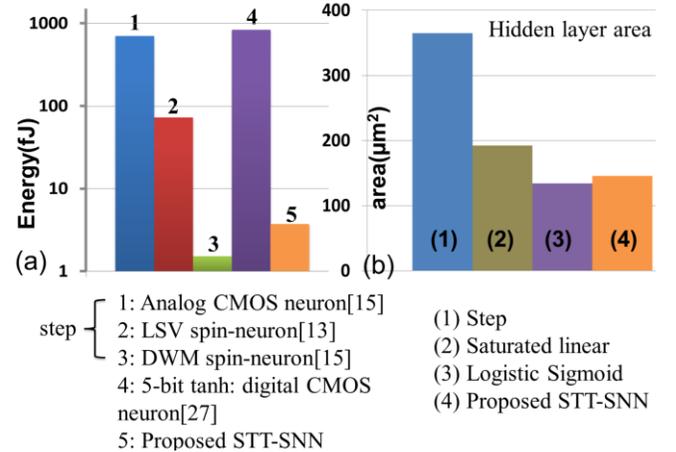

step
1: Analog CMOS neuron[15]
2: LSV spin-neuron[13]
3: DWM spin-neuron[15]
4: 5-bit tanh: digital CMOS neuron[27]
5: Proposed STT-SNN

(1) Step
(2) Saturated linear
(3) Logistic Sigmoid
(4) Proposed STT-SNN

Fig. 12 (a) Energy for different single neuron implementations, (b) hidden layer area based on different neuron transfer functions

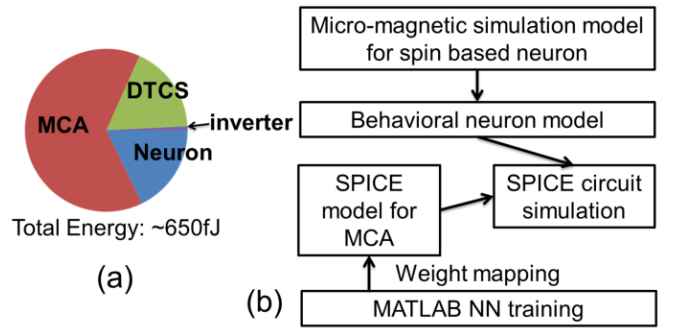

Fig. 13 (a) Energy analysis of the proposed ANN hardware for character recognition benchmark, (b) simulation framework used in this work distance between two memristors in the crossbar is ~300nm [26] and DTCS width is kept at 1µm. The hidden layer area using four different neuron transfer functions is shown in Fig. 12b. The hidden layers using soft-limiting neurons consume much smaller area because of less number of synapses and neurons. STT-SNN leads to ~2.5× lower hidden layer area compared to the hard-limiting step function neuron based ANN. The system level SPICE simulation of our proposed ANN hardware shows the total energy consumption for one alphabet recognition is ~650fJ (Fig. 13a), which is ~6.8× lower than that of the LSV neuron (step function) based ANN and more than two orders magnitude lower than the digital/analog ANN implementation for the same benchmark [13]. Note that, ANN training is performed offline and the programming of MCA-synapses is a one-time operation. Hence, the memristor programming energy is not included in our analysis.

A self-explanatory pictorial depiction of the simulation framework used in this work is given in Fig.13b. We used micro-magnetic simulation for the proposed STT-SNN and it was calibrated with experimental data from [9]. A compact behavioral model of STT-SNN was used in SPICE simulation. The ANN was trained offline using MATLAB's Neural Network toolbox [17], which generates the synapse weight matrix for the hidden and output layers from the given training data. The memristor conductance (1kΩ to 32kΩ, [14]) was programmed based on the synapse weight matrix in SPICE. In the system simulation, a Gaussian distributed random noise ($\sigma=5\%$) was added to each memristor conductance value to account for variations.

## VII. Conclusion

We propose a domain wall motion based spin-torque device that can efficiently implement a neuron with a soft-limiting non-linear transfer function, operating at ultra-low supply voltage and current. The spin based neuron device allows the peripheral circuits and memristor crossbar array synapses to also operate at very low voltages, thereby leading to ultra-low power consumption for the whole system. The proposed neurons are used to design artificial neural networks that show more than two orders of magnitude lower energy dissipation compared with analog and digital CMOS ANN implementations in 45nm CMOS technology and ~2.5× lower hidden layer area compared with hard-limiting neuron based ANNs. We believe that the proposed spin-transfer-torque based soft-limiting non-linear neurons along with MCA-synapses can be used to build energy efficient neuromorphic computing hardware for cognitive computing applications.

## Acknowledgment

This research was funded in part by the DARPA UPSIDE program, National Science Foundation under grant CCF-1320808, the Center for Spintronics: Materials, Interfaces, and Architecture (CSPIN), a StartNet Center funded by DoD and SRC, NSSEFF Fellows program, and the Semiconductor Research Corporation.